\documentclass[aps,prb,twocolumn,superscriptaddress, showpacs]{revtex4-1}

\usepackage{amsmath}
\usepackage{graphicx}
\begin{document}


\title{Coherent Control of Plasmonic Spectra using the Orbital Angular Momentum of Light}


\author{Aaron S. Rury}
\email[]{arury@caltech.edu}
\affiliation{Jet Propulsion Laboratory, California Institute of Technology, 4800 Oak Grove Dr., Pasadena, CA, 91109}
\affiliation{Applied Physics Program, University of Michigan, 450 Church St., Ann Arbor, MI 48109}

\date{\today}

\begin{abstract}
This study proposes a method to control the frequency-dependent scattering spectra from plasmonic spheres via the conservation of incident orbital angular momentum (OAM) in classical light scattering. By providing controllable distributions of OAM content, fractional vortex beams allow selective tailoring of Fano features present in coherent scattering processes. The applicability of this control methodology is briefly discussed in the context of plasmonic crystals that recent studies have shown possess modes described by a well-defined OAM content.    
\end{abstract}


\maketitle

Plasmonics has revolutionized optical physics by allowing researchers to couple electromagnetic radiation into material structures much shorter than a wavelength of light.\cite{MaierPlasmonics} By localizing electromagnetic excitations to interfaces between metals and dielectrics, surface plasmon polaritons provide probes of sub-micron environments as well as a pathway to sub-wavelength optical technologies. 

The asymmetric lineshape of some frequency-dependent spectra represents one of the more fascinating properties of plasmonic materials. Genet \emph{et al.} and Sarrazin \emph{et al.} simultaneously first proposed that these asymmetric features arise due to Fano processes in which participating electromagnetic modes interfere with one another in the scattered fields, analogous to the interference that gives rise to asymmetric lineshapes in autoionization processes discovered by Fano.\cite{Sarrazin,WoerdmanFano,Fano} Subsequent studies have found that these Fano structures appear in the spectra of many varieties of plasmonic materials including meta-materials based on plasmonic excitations.\cite{FanoNatureMaterials}  Plasmonic materials that display Fano spectral lineshapes have found use in applications ranging from nonlinear optical switching to high-resolution molecular spectroscopy to wavelength discrimination in nano-structures.\cite{Giant_Optical_Switching,Fano_molecular_spectroscopy,ControlFanoLineshapes} 

Despite the importance of interference in tailoring extinction, absorption and scattering spectra of plasmonic structures, active tuning of these spectral properties has remained elusive. To date, researchers have relied on changes in fabrication techniques and modulations of the dielectric environment surrounding metallic nano and micro-structures to change plasmonic spectra.\cite{THz_plasmonic_modulator,Atwater_plasmonic_modulator,ControlFanoLineshapes,ControllingFano} Despite the fidelity of these techniques, they are difficult to implement in real-time. For plasmonics to reach its full potential in many photonics applications which use easy, reliable and fast frequency dependent modulation of electromagnetic fields, this theoretical and technological gap must be filled. 

At the same time plasmonics began to change the landscape of optical materials, vortex beams emerged and provided researchers a new parameter for optical physics in the form of the orbital angular momentum (OAM) of light.\cite{Allen1992} Stemming from the azimuthal variation of the transverse profile of these light beams, researchers propose that OAM amends fundamental physics both in classical and quantum mechanical light-matter interactions.\cite{Alex,Babiker, Rury} However, to date the importance of OAM to plasmonics and other surface-localized electromagnetic excitations has yet to be fully elucidated.

This study explores the interaction of specific types of vortex beams with plasmonic structures whose modes possess well-defined angular momenta: metallic spherical particles. Using fractional vortex beams that possess a non-integer OAM content, this study shows that the asymmetric structures of coherent plasmonic scattering spectra change in response to changes in the distribution of different OAM states comprising the incident field. Based on these changes this letter proposes a model to coherently control the frequency-dependent plasmonic response in real-time with fractional vortex beams and the OAM of light. This proposal provides a method to actively control the position and magnitude of specific features in plasmonic spectra, thus advancing plasmonics in optical technologies ranging from material sensing and characterization to frequency discrimination and multiplexing. 


Recent research highlights the role of the conservation of angular momentum in modulating the classical electromagnetic scattering response of spherical structures.\cite{Rury} Using boundary conditions set by spherical symmetry, this earlier study found that the properties of the fields scattered from these structures depend sensitively on the angular momentum content of the incident field. Specifically, when the incident field carries both angular momentum associated with its transverse profile, OAM, and its polarization, multipole moments higher in order than the dipole moment become preferentially excited. Since plasmonic modes of spherical structures possess well-defined angular momenta, the conservation of angular momentum in scattering processes from these plasmonic states may also dictate phenomena that has been overlooked so far. 

\begin{figure*}[ht!]
\begin{center}
\centerline{\includegraphics[width=18 cm]{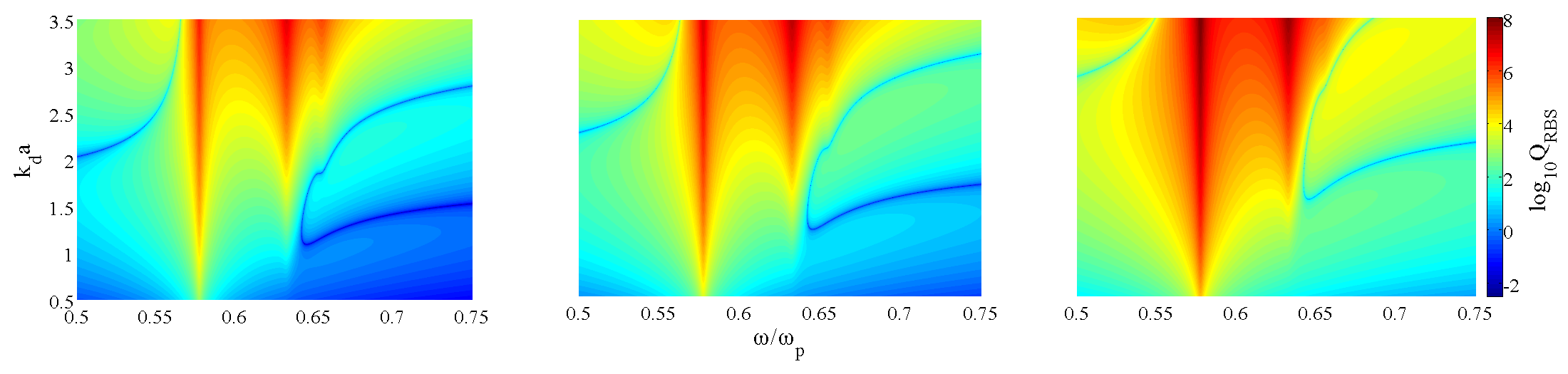}}
\caption{\label{FIG. 1}(color online) Two dimensional contour plots of the log of the coherent Radar Backscattering ($Q_{RBS}$) efficiency spectra of a metallic spherical particle with the optical properties of gold as a function of frequency ($\omega/\omega_p$) and the size parameter of the particle ($k_da$) for the cases of model fractional vortex beams ${L}_2$ (left), ${L}_{2*}$ (middle) and ${L}_0$ (right).}    
\end{center}
\end{figure*}

To control the distribution of fields that determine a material's plasmonic response, one must control the angular momentum states that contribute to the incident field. Fractional vortex beams possess a non-integer OAM content allowing precise control of a distribution of optical angular momentum states in a given light beam and have been fabricated in lab.\cite{Frac} To understand the role of the OAM of fractional vortex beams in this study, they are approximated as a linear superposition of purely azimuthal Laguerre-Gauss (LG) beams, each possessing a different azimuthal index, $\ell$, which denotes the amount of OAM content each state contributes to the incident field, as would be produced via etalon effects in vortex phase plates as shown previously.\cite{Rumala:13} That is, the transverse profile of an electric field of an incident fractional vortex beam with a center frequency $\omega$ focused to a waist $w_0$ and propagating in the $z$-direction is treated as,
\begin{align}
\vec{E}_{inc}=\sum_{\ell=\ell_{min}}^{\ell_{max}}\vec{E}_0g\left(\ell\right)\left(\frac{r\sqrt{2}}{w_0}\right)^\ell exp\left(\frac{-r^2}{w_0^2}\right)e^{\imath\ell\phi}e^{\imath k_d z},
\end{align}
where the factor $g\left(\ell\right)$ determines the weight with which a given OAM state contributes to the incident field, $\ell_{min}$ and $\ell_{max}$ represent the smallest and largest values of the azimuthal index that contribute to the distribution of OAM states of the field, respectively, and the vector amplitude, $\vec{E}_0$, describes a right circularly polarized state.

To understand the role of the conservation of angular momentum in modulating asymmetric features of plasmonic spectra, the field in Eq. (1) is incident on a metallic spherical particle of radius $a$ whose index of refraction relative to a surrounding dielectric medium is $N$. Based on the results of Ref. [15], in the limit of a small size parameter, the electric scattering coefficient due to a vortex beam carrying $\ell$ units of OAM is,
\begin{align}
\beta^{\ell,m}=\frac{E_{m\ell}}{\left(k_dw_0\right)^{\ell}}\frac{\rho^{(2m+\ell+1)}}{(2m+1)!!(2m-1)!!}\nonumber\\\times\frac{N^2\ell+(m+1)(N^2-1)}{1+m(N^2+1)}
\end{align}
where $m$ is the projection of the incident field's total angular momentum onto its direction of propagation, $\rho=k_da$ is the size parameter of the particle and $E_{m\ell}$ is related to the vector amplitude of the incident electric field and normalization of the multipole orders.  Eq. (2) clearly shows a term that combines the azimuthal index of the incident field, $\ell$, with the frequency dependent index of refraction of the particle, $N$ that does not appear in the case of an incident plane wave. In the case that the magnetic response of the particle can be ignored, these coefficients describe the scattered electric field through a standard definition.\cite{Jackson}


The frequency dependent information that determines the scattered field is contained within the electric scattering coefficient in Eq. (2) via the relative index of refraction, $N$, as defined above.
Using the Drude model under the assumption that the incident frequency is significantly larger than the material's interband transitions, the frequency-dependent index of refraction of the metallic particle relative to its dielectric surroundings becomes, 
\begin{align}
N^2(\omega)=1-\frac{\omega_p^2}{\omega^2}+\imath \frac{\omega_p^2\gamma}{\omega^3}
\end{align}
where $\omega_p$ is the plasma frequency of the metal and the $\gamma$ defines the decay rate of its electronic excitations. Inserting Eq. (3) into Eq. (2), we find that the frequency dependent electric scattering coefficient is,
\begin{widetext}
\begin{align}
\beta^{\ell,m}(\omega)=A_{m\ell}\left[\frac{\ell \frac{\omega}{\omega_p}\left(\frac{\omega^2}{\omega_p^2}+\frac{\gamma^2}{\omega_p^2}\right)-\left(\ell+m+1\right)\frac{\omega}{\omega_p}+\imath\left(\ell+m+1\right)\frac{\gamma}{\omega_p}}{\frac{\omega}{\omega_p}\left(2m+1\right)\left(\frac{\omega^2}{\omega_p^2}+\frac{\gamma^2}{\omega_p^2}\right)-m\frac{\omega}{\omega_p}+\imath m\frac{\gamma}{\omega_p}}\right],
\end{align}
\end{widetext}

where 

\begin{align}
A_{m\ell}=\frac{E_{m\ell}}{\left(k_dw_0\right)^{\ell}}\frac{q^{(2m+\ell+1)}}{(2m+1)!!(2m-1)!!}.
\end{align}
Therefore, to simulate the scattering spectra excited by the incident model fractional vortex beam in Eq. (1), one needs to use the appropriate coefficients and insert them into equations describing  coherent classical electromagnetic scattering.

\begin{figure*}[ht!]
\begin{center}
\centerline{\includegraphics[width=18 cm]{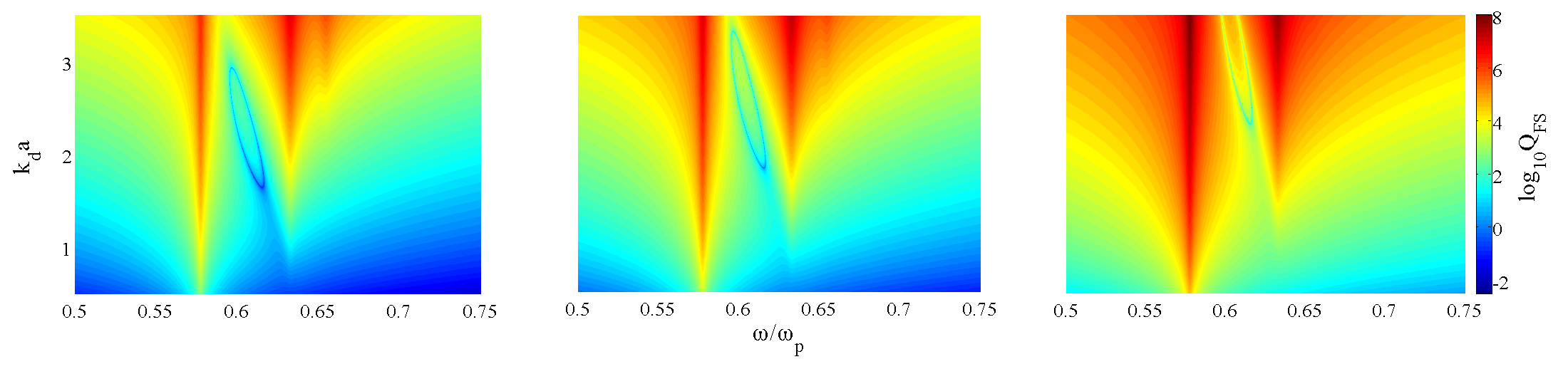}}
\caption{\label{FIG. 2}(color online) Two dimensional contour plots of the log of the coherent Forward Scattering ($Q_{FS}$) efficiency spectra of a metallic spherical particle with the optical properties of gold as a function of frequency ($\omega/\omega_p$) and the size parameter of the particle ($k_da$) for the cases of model fractional vortex beams ${L}_2$ (left), ${L}_{2^*}$ (middle) and ${L}_0$ (right).}    
\end{center}
\end{figure*} 

We treat Radar Backscattering (RBS) and Forward Scattering (FS) scattering processes since they necessitate a square of sum of scattering coefficients for simulation and thus contain coherent interference between different multipole orders. Explicitly, the efficiencies for these processes are \cite{FanoNatureMaterials},
\begin{subequations}
\begin{align}
Q_{RBS}=\frac{1}{\rho^2}\left|\sum_m(2m+1)(-1)^m\beta(\ell,m)\right|^2,\\
Q_{FS}=\frac{1}{\rho^2}\left|\sum_m(2m+1)\beta(\ell,m)\right|^2.
\end{align}
\end{subequations}

To simulate these scattering processes in the case of an incident fractional vortex beam as in Eq. (1), we use a standard Drude model explained above, set $\gamma/\omega_p=10^{-3}$ to correctly simulate plasmonic behavior of noble metals and use three different fractional vortex beams that possess three different distributions of angular momentum content in the $\ell=0$, $\ell=1$ and $\ell=2$ OAM states.\cite{bohren,Drude_constants_gold} We also assume a diffraction-limited spot size, $w_0$, such that $k_dw_0=\pi$. 


We denote the three representative fractional vortex beams by their dominant OAM contribution with the notation $L_n$. The first possesses the weighting factors $g(0)=\sqrt{0.01}$, $g(1)=\sqrt{0.04}$ and $g(2)=\sqrt{0.95}$ and is denoted $L_2$. The second beam possesses the weighting factors  $g(0)=\sqrt{0.05}$, $g(1)=\sqrt{0.1}$ and $g(2)=\sqrt{0.85}$ and is denoted $L_{2^*}$ since the $\ell=2$ state still dominates the distribution, but to a slightly lesser extent. The weighting factors $g(0)=\sqrt{0.5}$, $g(1)=\sqrt{0.25}$ and $g(2)=\sqrt{0.25}$ describe a vortex beam dominated by the $\ell=0$ state, denoted $L_0$. These parameters are meant as model vortex beams that allow one to see how weighting different OAM states in the incident field affect the observed spectra and are not exhaustive. One could choose other weighting parameters to excite different scattering spectra. The flexibility of these parameters in the incident field is the key to control of plasmonic excitation with fractional vortex beams.

We then assume that the lowest order contribution to the scattered field from each associated incident OAM state dominates the behavior of the scattered electric field and use the associated electric scattering coefficients as defined by Eqs. (2) and (3)  [$\beta(0,1)$, $\beta(1,2)$ and $\beta(2,3)$] in Eqs. (6a) and (6b). Figures (1) and (2) show two-dimensional (2D) contour spectra of each of the scattering efficiencies $Q_{RBS}$, and $Q_{FS}$ in log$_{10}$ units for each of the fractional vortex beams as functions of both the incident frequency scaled to the plasma frequency ($\omega/\omega_p$) and the size parameter of the particle ($k_da$). Figures (3a) and (3b) show a slice of each of the contour spectra for a size parameter $\rho\approx2.15$ as a function of frequency to compare the asymmetries of these spectra. These slices allow highlighting of the associated changes when the OAM of the incident field is changed. The structure of each set of spectra show significant dependence on the distribution of OAM states present in the incident field.

Each of the 2D spectra display distinct dips due to the interference between different multipole orders of the scattered field in the form of blue streaks and contours across each spectrum. The positions of these dips depend sensitively on the contribution of each mulitpole order to the scattered field. The 2D contour plots in Figure (2) and (3) show that the contribution of each mulitpole can be affected in two ways. On one hand, changes in the size parameter of the particle cause changes in the contribution of different multipole orders in the scattered fields since the particle's size determines the multipole expansion of its associated electronic charge density. In the cases studied here, the smallest particle sizes cause no interference in the observed spectra due to the dominance of the particle's dipole moment, Conversely, complex structures appear in the scattering spectra for both RBS and FS for larger size parameters due to the increased influence of higher order multipole orders of the particle's electrons. 

One the other hand, as one changes the distribution of OAM states that comprise the incident field, angular momentum conservation dictates which multipole orders are allowed to contribute to the scattered fields, thus changing the interference present in the scattered field. For instance, as the OAM content decreases from the $L_2$ beam to $L_0$, the blue ellipse between the electric dipole and quadrapole scattering orders in the 2D FS spectra moves from smaller to large size parameters and shifts to lower frequency. This change is seen in the 1D frequency dependent spectrum of FS in Figure (3b) as two dips moving more closely together until they are indistinguishable as the amount of OAM in the incident field is decreased. 

\begin{figure}
\centerline{\includegraphics[width=9 cm]{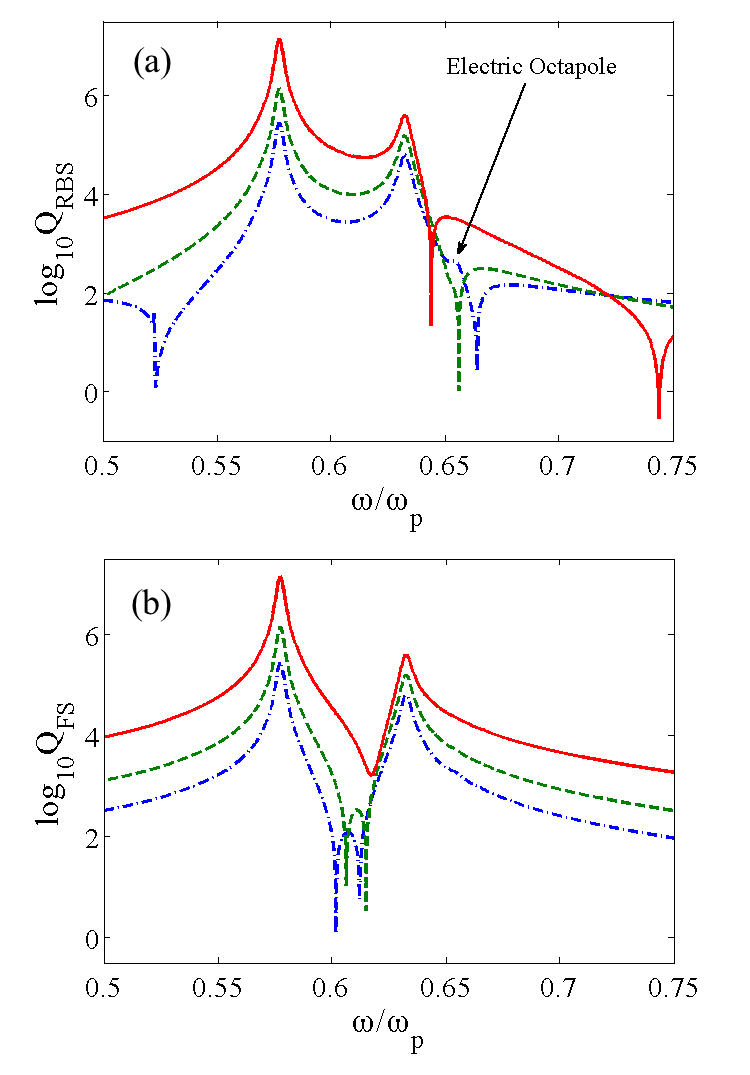}}
\caption{\label{FIG. 3}(color online) Frequency dependent slices of the 2D contour spectra for Radar Backscattering (a) and Forward Scattering (b) in the case of an incident $L_2$ beam (dash-dotted), $L_{2*}$ beam (dashed) and $L_0$ beam (solid).}
\end{figure}

Figure (3a) points out another, more conspicuous example of how the OAM density can cause noticeable and useful changes in coherent plasmonic scattering spectra. For this case, the region around the peak associated with the electric octapole plasmonic resonance is transformed into a dip in the spectrum as the incident beam changes from $L_2$ to $L_{2*}$. According to the simulations in this study, changing distribution of OAM states contributing to these two incident field changes the magnitude of the scattering efficiency by more than a factor of 100 around this spectral feature. In order to change the magnitude of the scattering efficiency by that amount using the size parameter or any other attribute of the material itself, one would need to fabricate a new sample of particles with different physical properties. Given the relative ease of changing OAM states using modern optical devices such as spatial light modulators and vortex phase plates, using an incident field to cause the same changes is highly advantageous for the further advancement of plasmonics in optical technologies.  

While spherical structures display plasmonic modes possessing well-defined angular momentum content, they are not the only material structures whose plasmonic properties are affected by angular momentum. Several groups have constructed plasmonic crystals whose collective modes lead to asymmetric Fano and Fano-like features in absorption and scattering spectra as well as chiral plasmonic materials that are capable of distinguishing angular momentum states of exciting light fields.\cite{Bullseye2008,Bullseye2011,chiralplasmonicsreview} Perhaps most interestingly in light of the current study, Gorodetski \emph{et al.} have shown that plasmonic crystals support the necessary angular momentum content to produce propagating beams that contain OAM via the interference between the multipolar moments of bulls-eye metallic structures.\cite{far-fieldOAM} However, all studies thus far have relied on precise fabrication techniques that necessitate substantial time scales to modulate the angular momentum-related plasmonic response of a material structure. Since the changes in the OAM states that contribute to an incident field do not rely on changes to the material structures as done in previous reports, the OAM-tailored plasmonic spectra modeled in this study may provide more reliable and reproducible control in plasmonics. 

Based on these considerations, I propose the use of fractional vortex beams and their tunable angular momentum content to control the coherent frequency-dependent response of a plasmonic crystal whose modes possess a well-defined angular momentum content in real time, such as that investigated by Gorodetski \emph{et al.}. By changing the distribution of OAM states that define an incident field, changes in interference mechanisms between participating multipolar fields may be able to tailor the frequencies at which strong or weak absorption is present and dramatically change the amount of light capable of passing through these structures. Such tailoring would provide avenues to new technologies in molecular detection and characterization, optical switching as well as frequency discrimination and multiplexing based on coherent plasomic responses. 

\section{acknowledgments}
This work was supported by the Defense Threat Reduction Agency-Joint Science and Technology Office for Chemical and Biological Defense (grant HDTRA1-09-1-0005).


\begin{thebibliography}{24}%
\makeatletter
\providecommand \@ifxundefined [1]{%
 \@ifx{#1\undefined}
}%
\providecommand \@ifnum [1]{%
 \ifnum #1\expandafter \@firstoftwo
 \else \expandafter \@secondoftwo
 \fi
}%
\providecommand \@ifx [1]{%
 \ifx #1\expandafter \@firstoftwo
 \else \expandafter \@secondoftwo
 \fi
}%
\providecommand \natexlab [1]{#1}%
\providecommand \enquote  [1]{``#1''}%
\providecommand \bibnamefont  [1]{#1}%
\providecommand \bibfnamefont [1]{#1}%
\providecommand \citenamefont [1]{#1}%
\providecommand \href@noop [0]{\@secondoftwo}%
\providecommand \href [0]{\begingroup \@sanitize@url \@href}%
\providecommand \@href[1]{\@@startlink{#1}\@@href}%
\providecommand \@@href[1]{\endgroup#1\@@endlink}%
\providecommand \@sanitize@url [0]{\catcode `\\12\catcode `\$12\catcode
  `\&12\catcode `\#12\catcode `\^12\catcode `\_12\catcode `\%12\relax}%
\providecommand \@@startlink[1]{}%
\providecommand \@@endlink[0]{}%
\providecommand \url  [0]{\begingroup\@sanitize@url \@url }%
\providecommand \@url [1]{\endgroup\@href {#1}{\urlprefix }}%
\providecommand \urlprefix  [0]{URL }%
\providecommand \Eprint [0]{\href }%
\providecommand \doibase [0]{http://dx.doi.org/}%
\providecommand \selectlanguage [0]{\@gobble}%
\providecommand \bibinfo  [0]{\@secondoftwo}%
\providecommand \bibfield  [0]{\@secondoftwo}%
\providecommand \translation [1]{[#1]}%
\providecommand \BibitemOpen [0]{}%
\providecommand \bibitemStop [0]{}%
\providecommand \bibitemNoStop [0]{.\EOS\space}%
\providecommand \EOS [0]{\spacefactor3000\relax}%
\providecommand \BibitemShut  [1]{\csname bibitem#1\endcsname}%
\let\auto@bib@innerbib\@empty
\bibitem [{\citenamefont {Maier}\ \emph {et~al.}(2001)\citenamefont {Maier},
  \citenamefont {Brongersma}, \citenamefont {Kik}, \citenamefont {Meltzer},
  \citenamefont {Requicha},\ and\ \citenamefont {Atwater}}]{MaierPlasmonics}%
  \BibitemOpen
  \bibfield  {author} {\bibinfo {author} {\bibfnamefont {S.}~\bibnamefont
  {Maier}}, \bibinfo {author} {\bibfnamefont {M.}~\bibnamefont {Brongersma}},
  \bibinfo {author} {\bibfnamefont {P.}~\bibnamefont {Kik}}, \bibinfo {author}
  {\bibfnamefont {S.}~\bibnamefont {Meltzer}}, \bibinfo {author} {\bibfnamefont
  {A.}~\bibnamefont {Requicha}}, \ and\ \bibinfo {author} {\bibfnamefont
  {H.}~\bibnamefont {Atwater}},\ }\href {\doibase
  {10.1002/1521-4095(200110)13:19<1501::AID-ADMA1501>3.0.CO;2-Z}} {\bibfield
  {journal} {\bibinfo  {journal} {{Advanced Materials}}\ }\textbf {\bibinfo
  {volume} {{13}}},\ \bibinfo {pages} {{1501}} (\bibinfo {year}
  {{2001}})}\BibitemShut {NoStop}%
\bibitem [{\citenamefont {{Sarrazin, Micha\"el and Vigneron, Jean-Pol and
  Vigoureux, Jean-Marie}}(2003)}]{Sarrazin}%
  \BibitemOpen
  \bibfield  {author} {\bibinfo {author} {\bibnamefont {{Sarrazin, Micha\"el
  and Vigneron, Jean-Pol and Vigoureux, Jean-Marie}}},\ }\href@noop {}
  {\bibfield  {journal} {\bibinfo  {journal} {{Phys. Rev. B}}\ }\textbf
  {\bibinfo {volume} {{67}}},\ \bibinfo {pages} {{085415}} (\bibinfo {year}
  {{2003}})}\BibitemShut {NoStop}%
\bibitem [{\citenamefont {Genet}\ \emph {et~al.}(2003)\citenamefont {Genet},
  \citenamefont {van Exter},\ and\ \citenamefont {Woerdman}}]{WoerdmanFano}%
  \BibitemOpen
  \bibfield  {author} {\bibinfo {author} {\bibfnamefont {C.}~\bibnamefont
  {Genet}}, \bibinfo {author} {\bibfnamefont {M.}~\bibnamefont {van Exter}}, \
  and\ \bibinfo {author} {\bibfnamefont {J.}~\bibnamefont {Woerdman}},\ }\href
  {\doibase {10.1016/j.optcom.2003.07.037}} {\bibfield  {journal} {\bibinfo
  {journal} {{Opt. Comm.}}\ }\textbf {\bibinfo {volume} {{225}}},\ \bibinfo
  {pages} {{331} (\bibinfo {year} {{2003}})}\BibitemShut {NoStop}%
\bibitem [{\citenamefont {{Fano, U}}(1961)}]{Fano}%
  \BibitemOpen
  \bibfield  {author} {\bibinfo {author} {\bibnamefont {{Fano, U}}},\
  }\href@noop {} {\bibfield  {journal} {\bibinfo  {journal} {{Phys. Rev.}}\
  }\textbf {\bibinfo {volume} {{124}}},\ \bibinfo {pages} {{1866} (\bibinfo
  {year} {{1961}})}\BibitemShut {NoStop}%
\bibitem [{\citenamefont {Luk'yanchuk}\ \emph {et~al.}(2010)\citenamefont
  {Luk'yanchuk}, \citenamefont {Zheludev}, \citenamefont {Maier}, \citenamefont
  {Halas}, \citenamefont {Nordlander}, \citenamefont {Giessen},\ and\
  \citenamefont {Chong}}]{FanoNatureMaterials}%
  \BibitemOpen
  \bibfield  {author} {\bibinfo {author} {\bibfnamefont {B.}~\bibnamefont
  {Luk'yanchuk}}, \bibinfo {author} {\bibfnamefont {N.~I.}\ \bibnamefont
  {Zheludev}}, \bibinfo {author} {\bibfnamefont {S.~A.}\ \bibnamefont {Maier}},
  \bibinfo {author} {\bibfnamefont {N.~J.}\ \bibnamefont {Halas}}, \bibinfo
  {author} {\bibfnamefont {P.}~\bibnamefont {Nordlander}}, \bibinfo {author}
  {\bibfnamefont {H.}~\bibnamefont {Giessen}}, \ and\ \bibinfo {author}
  {\bibfnamefont {C.~T.}\ \bibnamefont {Chong}},\ }\href {\doibase
  {10.1038/NMAT2810}} {\bibfield  {journal} {\bibinfo  {journal} {{Nature
  Mat.}}\ }\textbf {\bibinfo {volume} {{9}}},\ \bibinfo {pages} {{707}
  (\bibinfo {year} {{2010}})}\BibitemShut {NoStop}%
\bibitem [{\citenamefont {Argyropoulos}\ \emph {et~al.}(2012)\citenamefont
  {Argyropoulos}, \citenamefont {Chen}, \citenamefont {Monticone},
  \citenamefont {D'Aguanno},\ and\ \citenamefont
  {Al{\`u}}}]{Giant_Optical_Switching}%
  \BibitemOpen
  \bibfield  {author} {\bibinfo {author} {\bibfnamefont {C.}~\bibnamefont
  {Argyropoulos}}, \bibinfo {author} {\bibfnamefont {P.-Y.}\ \bibnamefont
  {Chen}}, \bibinfo {author} {\bibfnamefont {F.}~\bibnamefont {Monticone}},
  \bibinfo {author} {\bibfnamefont {G.}~\bibnamefont {D'Aguanno}}, \ and\
  \bibinfo {author} {\bibfnamefont {A.}~\bibnamefont {Al{\`u}}},\ }\href
  {http://link.aps.org/doi/10.1103/PhysRevLett.108.263905} {\bibfield
  {journal} {\bibinfo  {journal} {{Phys. Rev. Lett.}}\ }\textbf {\bibinfo
  {volume} {{108}}},\ \bibinfo {pages} {263905} (\bibinfo {year}
  {{2012}})}\BibitemShut {NoStop}%
\bibitem [{\citenamefont {Wu}\ \emph {et~al.}(2012)\citenamefont {Wu},
  \citenamefont {Khanikaev}, \citenamefont {Adato}, \citenamefont {Arju},
  \citenamefont {Yanik}, \citenamefont {Altug},\ and\ \citenamefont
  {Shvets}}]{Fano_molecular_spectroscopy}%
  \BibitemOpen
  \bibfield  {author} {\bibinfo {author} {\bibfnamefont {C.}~\bibnamefont
  {Wu}}, \bibinfo {author} {\bibfnamefont {A.~B.}\ \bibnamefont {Khanikaev}},
  \bibinfo {author} {\bibfnamefont {R.}~\bibnamefont {Adato}}, \bibinfo
  {author} {\bibfnamefont {N.}~\bibnamefont {Arju}}, \bibinfo {author}
  {\bibfnamefont {A.~A.}\ \bibnamefont {Yanik}}, \bibinfo {author}
  {\bibfnamefont {H.}~\bibnamefont {Altug}}, \ and\ \bibinfo {author}
  {\bibfnamefont {G.}~\bibnamefont {Shvets}},\ }\href
  {http://dx.doi.org/10.1038/nmat3161} {\bibfield  {journal} {\bibinfo
  {journal} {{Nature Mat.}}\ }\textbf {\bibinfo {volume} {{11}}},\ \bibinfo
  {pages} {69} (\bibinfo {year} {{2012}})}\BibitemShut {NoStop}%
\bibitem [{\citenamefont {Spinelli}\ \emph {et~al.}(2011)\citenamefont
  {Spinelli}, \citenamefont {van Lare}, \citenamefont {Verhagen},\ and\
  \citenamefont {Polman}}]{ControlFanoLineshapes}%
  \BibitemOpen
  \bibfield  {author} {\bibinfo {author} {\bibfnamefont {P.}~\bibnamefont
  {Spinelli}}, \bibinfo {author} {\bibfnamefont {C.}~\bibnamefont {van Lare}},
  \bibinfo {author} {\bibfnamefont {E.}~\bibnamefont {Verhagen}}, \ and\
  \bibinfo {author} {\bibfnamefont {A.}~\bibnamefont {Polman}},\ }\href@noop {}
  {\bibfield  {journal} {\bibinfo  {journal} {{Opt. Express}}\ }\textbf
  {\bibinfo {volume} {{19}}},\ \bibinfo {pages} {{A303} (\bibinfo {year}
  {{2011}})}\BibitemShut {NoStop}%
\bibitem [{\citenamefont {Dintinger}\ \emph {et~al.}(2006)\citenamefont
  {Dintinger}, \citenamefont {Robel}, \citenamefont {Kamat}, \citenamefont
  {Genet},\ and\ \citenamefont {Ebbesen}}]{THz_plasmonic_modulator}%
  \BibitemOpen
  \bibfield  {author} {\bibinfo {author} {\bibfnamefont {J.}~\bibnamefont
  {Dintinger}}, \bibinfo {author} {\bibfnamefont {I.}~\bibnamefont {Robel}},
  \bibinfo {author} {\bibfnamefont {P.~V.}\ \bibnamefont {Kamat}}, \bibinfo
  {author} {\bibfnamefont {C.}~\bibnamefont {Genet}}, \ and\ \bibinfo {author}
  {\bibfnamefont {T.~W.}\ \bibnamefont {Ebbesen}},\ }\href {\doibase
  {10.1002/adma.200600366}} {\bibfield  {journal} {\bibinfo  {journal}
  {{Advanced Materials}}\ }\textbf {\bibinfo {volume} {{18}}},\ \bibinfo
  {pages} {{1645}} (\bibinfo {year} {{2006}})}\BibitemShut {NoStop}%
\bibitem [{\citenamefont {Pacifici}\ \emph {et~al.}(2007)\citenamefont
  {Pacifici}, \citenamefont {Lezec},\ and\ \citenamefont
  {Atwater}}]{Atwater_plasmonic_modulator}%
  \BibitemOpen
  \bibfield  {author} {\bibinfo {author} {\bibfnamefont {D.}~\bibnamefont
  {Pacifici}}, \bibinfo {author} {\bibfnamefont {H.~J.}\ \bibnamefont {Lezec}},
  \ and\ \bibinfo {author} {\bibfnamefont {H.~A.}\ \bibnamefont {Atwater}},\
  }\href {\doibase {10.1038/nphoton.2007.95}} {\bibfield  {journal} {\bibinfo
  {journal} {{Nature Photonics}}\ }\textbf {\bibinfo {volume} {{1}}},\ \bibinfo
  {pages} {{402} (\bibinfo {year} {{2007}})}\BibitemShut {NoStop}%
\bibitem [{\citenamefont {Christ}\ \emph {et~al.}(2007)\citenamefont {Christ},
  \citenamefont {Ekinci}, \citenamefont {Solak}, \citenamefont {Gippius},
  \citenamefont {Tikhodeev},\ and\ \citenamefont {Martin}}]{ControllingFano}%
  \BibitemOpen
  \bibfield  {author} {\bibinfo {author} {\bibfnamefont {A.}~\bibnamefont
  {Christ}}, \bibinfo {author} {\bibfnamefont {Y.}~\bibnamefont {Ekinci}},
  \bibinfo {author} {\bibfnamefont {H.~H.}\ \bibnamefont {Solak}}, \bibinfo
  {author} {\bibfnamefont {N.~A.}\ \bibnamefont {Gippius}}, \bibinfo {author}
  {\bibfnamefont {S.~G.}\ \bibnamefont {Tikhodeev}}, \ and\ \bibinfo {author}
  {\bibfnamefont {O.~J.~F.}\ \bibnamefont {Martin}},\ }\href {\doibase
  {10.1103/PhysRevB.76.201405}} {\bibfield  {journal} {\bibinfo  {journal}
  {{Phys. Rev. B}}\ }\textbf {\bibinfo {volume} {{76}}} (\bibinfo {year}
  {{2007}}),\ {10.1103/PhysRevB.76.201405}}\BibitemShut {NoStop}%
\bibitem [{\citenamefont {Allen}\ \emph {et~al.}(1992)\citenamefont {Allen},
  \citenamefont {Beijersbergen}, \citenamefont {Spreeuw},\ and\ \citenamefont
  {Woerdman}}]{Allen1992}%
  \BibitemOpen
  \bibfield  {author} {\bibinfo {author} {\bibfnamefont {L.}~\bibnamefont
  {Allen}}, \bibinfo {author} {\bibfnamefont {M.}~\bibnamefont
  {Beijersbergen}}, \bibinfo {author} {\bibfnamefont {R.}~\bibnamefont
  {Spreeuw}}, \ and\ \bibinfo {author} {\bibfnamefont {J.}~\bibnamefont
  {Woerdman}},\ }\href {\doibase {10.1103/PhysRevA.45.8185}} {\bibfield
  {journal} {\bibinfo  {journal} {{Phys. Rev. A}}\ }\textbf {\bibinfo {volume}
  {{45}}},\ \bibinfo {pages} {{8185} (\bibinfo {year} {{1992}})}\BibitemShut
  {NoStop}%
\bibitem [{\citenamefont {Alexandrescu}\ \emph {et~al.}(2006)\citenamefont
  {Alexandrescu}, \citenamefont {Cojoc},\ and\ \citenamefont
  {Fabrizio}}]{Alex}%
  \BibitemOpen
  \bibfield  {author} {\bibinfo {author} {\bibfnamefont {A.}~\bibnamefont
  {Alexandrescu}}, \bibinfo {author} {\bibfnamefont {D.}~\bibnamefont {Cojoc}},
  \ and\ \bibinfo {author} {\bibfnamefont {E.~D.}\ \bibnamefont {Fabrizio}},\
  }\href {\doibase 10.1103/PhysRevLett.96.243001} {\bibfield  {journal}
  {\bibinfo  {journal} {Phys. Rev. Lett.}\ }\textbf {\bibinfo {volume} {96}},\
  \bibinfo {pages} {243001} (\bibinfo {year} {2006})}\BibitemShut {NoStop}%
\bibitem [{\citenamefont {Babiker}\ \emph {et~al.}(2002)\citenamefont
  {Babiker}, \citenamefont {Bennett}, \citenamefont {Andrews},\ and\
  \citenamefont {Romero}}]{Babiker}%
  \BibitemOpen
  \bibfield  {author} {\bibinfo {author} {\bibfnamefont {M.}~\bibnamefont
  {Babiker}}, \bibinfo {author} {\bibfnamefont {C.}~\bibnamefont {Bennett}},
  \bibinfo {author} {\bibfnamefont {D.}~\bibnamefont {Andrews}}, \ and\
  \bibinfo {author} {\bibfnamefont {L.}~\bibnamefont {Romero}},\ }\href
  {\doibase {10.1103/PhysRevLett.89.143601}} {\bibfield  {journal} {\bibinfo
  {journal} {{Phys. Rev. Lett.}}\ }\textbf {\bibinfo {volume} {{89}}} (\bibinfo
  {year} {{2002}}),\ {10.1103/PhysRevLett.89.143601}}\BibitemShut {NoStop}%
\bibitem [{\citenamefont {Rury}\ and\ \citenamefont {Freeling}(2012)}]{Rury}%
  \BibitemOpen
  \bibfield  {author} {\bibinfo {author} {\bibfnamefont {A.~S.}\ \bibnamefont
  {Rury}}\ and\ \bibinfo {author} {\bibfnamefont {R.}~\bibnamefont
  {Freeling}},\ }\href {\doibase {10.1103/PhysRevA.86.053830}} {\bibfield
  {journal} {\bibinfo  {journal} {{Phys. Rev. A}}\ }\textbf {\bibinfo {volume}
  {{86}}} (\bibinfo {year} {{2012}}),\
  {10.1103/PhysRevA.86.053830}}\BibitemShut {NoStop}%
\bibitem [{\citenamefont {Goette}\ \emph {et~al.}(2008)\citenamefont {Goette},
  \citenamefont {O'Holleran}, \citenamefont {Preece}, \citenamefont
  {Flossmann}, \citenamefont {Franke-Arnold}, \citenamefont {Barnett},\ and\
  \citenamefont {Padgett}}]{Frac}%
  \BibitemOpen
  \bibfield  {author} {\bibinfo {author} {\bibfnamefont {J.~B.}\ \bibnamefont
  {Goette}}, \bibinfo {author} {\bibfnamefont {K.}~\bibnamefont {O'Holleran}},
  \bibinfo {author} {\bibfnamefont {D.}~\bibnamefont {Preece}}, \bibinfo
  {author} {\bibfnamefont {F.}~\bibnamefont {Flossmann}}, \bibinfo {author}
  {\bibfnamefont {S.}~\bibnamefont {Franke-Arnold}}, \bibinfo {author}
  {\bibfnamefont {S.~M.}\ \bibnamefont {Barnett}}, \ and\ \bibinfo {author}
  {\bibfnamefont {M.~J.}\ \bibnamefont {Padgett}},\ }\href {\doibase
  {10.1364/OE.16.000993}} {\bibfield  {journal} {\bibinfo  {journal} {{Opt.
  Express}}\ }\textbf {\bibinfo {volume} {{16}}},\ \bibinfo {pages} {{993}
  (\bibinfo {year} {{2008}})}\BibitemShut {NoStop}%
\bibitem [{\citenamefont {Rumala}\ and\ \citenamefont
  {Leanhardt}(2013)}]{Rumala:13}%
  \BibitemOpen
  \bibfield  {author} {\bibinfo {author} {\bibfnamefont {Y.~S.}\ \bibnamefont
  {Rumala}}\ and\ \bibinfo {author} {\bibfnamefont {A.~E.}\ \bibnamefont
  {Leanhardt}},\ }\href {\doibase 10.1364/JOSAB.30.000615} {\bibfield
  {journal} {\bibinfo  {journal} {J. Opt. Soc. Am. B}\ }\textbf {\bibinfo
  {volume} {30}},\ \bibinfo {pages} {615} (\bibinfo {year} {2013})}\BibitemShut
  {NoStop}%
\bibitem [{\citenamefont {Jackson}(1999)}]{Jackson}%
  \BibitemOpen
  \bibfield  {author} {\bibinfo {author} {\bibfnamefont {J.}~\bibnamefont
  {Jackson}},\ }\href {http://books.google.com/books?id=U3LBQgAACAAJ} {\emph
  {\bibinfo {title} {Classical Electrodynamics}}}\ (\bibinfo  {publisher}
  {Wiley},\ \bibinfo {year} {1999})\BibitemShut {NoStop}%
\bibitem [{\citenamefont {Bohren}\ and\ \citenamefont
  {Huffman}(1983)}]{bohren}%
  \BibitemOpen
  \bibfield  {author} {\bibinfo {author} {\bibfnamefont {C.}~\bibnamefont
  {Bohren}}\ and\ \bibinfo {author} {\bibfnamefont {D.}~\bibnamefont
  {Huffman}},\ }\href {http://books.google.com/books?id=R5IpAQAAMAAJ} {\emph
  {\bibinfo {title} {Absorption and scattering of light by small particles}}},\
  Wiley science paperback series\ (\bibinfo  {publisher} {Wiley},\ \bibinfo
  {year} {1983})\BibitemShut {NoStop}%
\bibitem [{\citenamefont {{Murata, Ken-ichiro and Tanaka,
  Hajime}}(2010)}]{Drude_constants_gold}%
  \BibitemOpen
  \bibfield  {author} {\bibinfo {author} {\bibnamefont {{Murata, Ken-ichiro and
  Tanaka, Hajime}}},\ }\href {http://dx.doi.org/10.1038/ncomms1015} {\bibfield
  {journal} {\bibinfo  {journal} {{Nature Commun.}}\ }\textbf {\bibinfo
  {volume} {1}} (\bibinfo {year} {2010})}\BibitemShut {NoStop}%
\bibitem [{\citenamefont {Drezet}\ \emph {et~al.}(2008)\citenamefont {Drezet},
  \citenamefont {Genet},\ and\ \citenamefont {Ebbesen}}]{Bullseye2008}%
  \BibitemOpen
  \bibfield  {author} {\bibinfo {author} {\bibfnamefont {A.}~\bibnamefont
  {Drezet}}, \bibinfo {author} {\bibfnamefont {C.}~\bibnamefont {Genet}}, \
  and\ \bibinfo {author} {\bibfnamefont {T.~W.}\ \bibnamefont {Ebbesen}},\
  }\href {\doibase {10.1103/PhysRevLett.101.043902}} {\bibfield  {journal}
  {\bibinfo  {journal} {{Phys. Rev. Lett.}}\ }\textbf {\bibinfo {volume}
  {{101}}} (\bibinfo {year} {{2008}}),\
  {10.1103/PhysRevLett.101.043902}}\BibitemShut {NoStop}%
\bibitem [{\citenamefont {Carretero-Palacios}\ \emph
  {et~al.}(2011)\citenamefont {Carretero-Palacios}, \citenamefont {Mahboub},
  \citenamefont {Garcia-Vidal}, \citenamefont {Martin-Moreno}, \citenamefont
  {Rodrigo}, \citenamefont {Genet},\ and\ \citenamefont
  {Ebbesen}}]{Bullseye2011}%
  \BibitemOpen
  \bibfield  {author} {\bibinfo {author} {\bibfnamefont {S.}~\bibnamefont
  {Carretero-Palacios}}, \bibinfo {author} {\bibfnamefont {O.}~\bibnamefont
  {Mahboub}}, \bibinfo {author} {\bibfnamefont {F.~J.}\ \bibnamefont
  {Garcia-Vidal}}, \bibinfo {author} {\bibfnamefont {L.}~\bibnamefont
  {Martin-Moreno}}, \bibinfo {author} {\bibfnamefont {S.~G.}\ \bibnamefont
  {Rodrigo}}, \bibinfo {author} {\bibfnamefont {C.}~\bibnamefont {Genet}}, \
  and\ \bibinfo {author} {\bibfnamefont {T.~W.}\ \bibnamefont {Ebbesen}},\
  }\href@noop {} {\bibfield  {journal} {\bibinfo  {journal} {{Optics Express}}\
  }\textbf {\bibinfo {volume} {{19}}},\ \bibinfo {pages} {{10429} (\bibinfo
  {year} {{2011}})}\BibitemShut {NoStop}%
\bibitem [{\citenamefont {Guerrero-Martinez}\ \emph {et~al.}(2011)\citenamefont
  {Guerrero-Martinez}, \citenamefont {Lorenzo Alonso-Gomez}, \citenamefont
  {Auguie}, \citenamefont {Magdalena~Cid},\ and\ \citenamefont
  {Liz-Marzan}}]{chiralplasmonicsreview}%
  \BibitemOpen
  \bibfield  {author} {\bibinfo {author} {\bibfnamefont {A.}~\bibnamefont
  {Guerrero-Martinez}}, \bibinfo {author} {\bibfnamefont {J.}~\bibnamefont
  {Lorenzo Alonso-Gomez}}, \bibinfo {author} {\bibfnamefont {B.}~\bibnamefont
  {Auguie}}, \bibinfo {author} {\bibfnamefont {M.}~\bibnamefont
  {Magdalena~Cid}}, \ and\ \bibinfo {author} {\bibfnamefont {L.~M.}\
  \bibnamefont {Liz-Marzan}},\ }\href {\doibase {10.1016/j.nantod.2011.06.003}}
  {\bibfield  {journal} {\bibinfo  {journal} {{Nano Today}}\ }\textbf {\bibinfo
  {volume} {{6}}},\ \bibinfo {pages} {{381} (\bibinfo {year}
  {{2011}})}\BibitemShut {NoStop}%
\bibitem [{\citenamefont {Gorodetski}\ \emph {et~al.}(2013)\citenamefont
  {Gorodetski}, \citenamefont {Drezet}, \citenamefont {Genet},\ and\
  \citenamefont {Ebbesen}}]{far-fieldOAM}%
  \BibitemOpen
  \bibfield  {author} {\bibinfo {author} {\bibfnamefont {Y.}~\bibnamefont
  {Gorodetski}}, \bibinfo {author} {\bibfnamefont {A.}~\bibnamefont {Drezet}},
  \bibinfo {author} {\bibfnamefont {C.}~\bibnamefont {Genet}}, \ and\ \bibinfo
  {author} {\bibfnamefont {T.~W.}\ \bibnamefont {Ebbesen}},\ }\href@noop {}
  {\bibfield  {journal} {\bibinfo  {journal} {{Phys. Rev. Lett.}}\ }\textbf
  {\bibinfo {volume} {{110}}},\ \bibinfo {pages} {{203906}} (\bibinfo {year}
  {{2013}})}\BibitemShut {NoStop}}}}}}}}}}
\end{thebibliography}
\end{document}